\documentclass[prd,twocolumn,showpacs,preprintnumbers,nofootinbib]{revtex4}
\usepackage{amsfonts,amsmath,amssymb}
\usepackage{graphicx}
\usepackage{color}
\usepackage{natbib}
\usepackage[plainpages=false, colorlinks=true, anchorcolor=blue, linkcolor=blue, citecolor=blue, bookmarks=false]{hyperref}
\pdfoutput=1

\usepackage{natbib}

\begin{document}

\newcommand{\apjl}{Astrophys. J. Lett.}
\newcommand{\apjs}{Astrophys. J. Suppl. Ser.}
\newcommand{\aap}{Astron. \& Astrophys.}
\newcommand{\aj}{Astron. J.}
\newcommand{\araa}{Ann. Rev. Astron. Astrophys. } 
\newcommand{\mnras}{Mon. Not. R. Astron. Soc.}
\newcommand{\jcap}{JCAP}
\newcommand{\pasj}{PASJ}
\newcommand{\pasa}{Pub. Astro. Soc. Aust.}
\newcommand{\physrep}{Phys. Rep.}

\title{Limit on graviton mass   from galaxy cluster Abell 1689}
\author{Shantanu  \surname{Desai}$^1$} \altaffiliation{E-mail: shntn05@gmail.com}

\affiliation{$^{1}$Department of Physics, Indian Institute of Technology, Hyderabad, Telangana-502285, India}

\begin{abstract}
To date, the only limit  on graviton mass using galaxy clusters was obtained by Goldhaber and Nieto in 1974,  using the fact that the orbits of galaxy clusters are bound and closed, and  extend up to 580 kpc. From positing that only a Newtonian potential gives rise to such stable bound orbits, a limit on the graviton mass  $m_g<1.1 \times 10^{-29}$ eV was obtained~\cite{Goldhaber74}. Recently, it has been shown that one can obtain closed bound orbits for Yukawa potential~\cite{Mukherjee}, thus invalidating the main \emph{ansatz} used in Ref.~\cite{Goldhaber74} to obtain the graviton mass bound. In order to  obtain a revised estimate using galaxy clusters, we use dynamical mass models of the Abell 1689 (A1689) galaxy cluster to check their compatibility  with a Yukawa gravitational potential. We use the   mass models for  the gas, dark matter, and galaxies for  A1689 from Refs.~\cite{Nieu,Hodson}, who used this cluster to test various alternate gravity theories, which dispense with the need for dark matter.  We quantify  the deviations in the  acceleration profile using these mass models assuming  a Yukawa potential and that obtained  assuming   a Newtonian potential by calculating the  $\chi^2$ residuals between the two profiles. Our estimated bound  on the graviton mass ($m_g$) is thereby given by,  $m_g < 1.37 \times 10^{-29}$ eV or in terms of  the graviton Compton wavelength of, $\lambda_g>9.1 \times 10^{19}$ km at 90\% confidence level.
\end{abstract}

\maketitle

\section{Introduction}

A century after its inception, General relativity (GR) passes all  observational tests at solar system and  binary pulsar length scales  with flying colors~\cite{Will, Turyshev,Stairs}. The recent direct detection of gravitational waves has confirmed the validity of general relativity in the dynamical strong-field regime~\cite{Yunes16}.
Despite this, a whole slew of modified theories of gravity have been explored, ever since the equations of GR were first written more than a century ago. Most  of the recent resurgence in  studying and proposing the plethora of  modified theories of gravity has been driven by the need to  address problems in Cosmology such as Dark matter, Dark Energy, Inflation, and Baryogenesis~\cite{Ferreira,Koyama,Khoury,Woodard06,Schmidt,Martin,Yunes,Poplawski,Bamba,Bamba2}, which cannot be explained using GR and the Standard model of particle physics. Independent of cosmological problems, a number of alternatives have also been extensively proposed  to resolve conceptual problems in classical GR at the interface of fundamental Physics, such as Big Bang singularity~\cite{Hawking}, arrow of time~\cite{Ellis,Padmanabhan}, or  the quantization of gravity~\cite{Carlip,Ashtekar,Woodard09}.  An updated summary of almost all the modified theories of gravity can be found in monographs such as~\cite{Will93} and also in recent  reviews~\cite{Will15,Baker,Smoot} and references therein.

One such modification to GR, namely massive gravity in which a graviton is endowed with  non-zero mass dates back to more than 70 years. The first ever theory  of massive gravity in the perturbative limit was proposed by Pauli and Fierz~\cite{Pauli}. However, this theory does not reproduce the GR result, in the limit when the  graviton mass goes to zero, usually referred to in the literature as the  vDVZ discontinuity~\cite{Veltman,Zakharov}. However, Vainshtein showed that this discontinuity is due to how the gravitational degrees of freedom are treated during the linearization procedure and can be fixed in a non-linear version of massive gravity~\cite{Vainshtein}. Bouleware and Deser then showed that this non-linear version  has a ghost~\cite{Deser}. Therefore, the field of massive gravity theories lay dormant because of these conceptual problems. However in the last decade, the Bouleware-Deser ghost problem  has been solved, leading to a resurgence of interest in these massive gravity  theories~\cite{Derham11,Rosen,Kurt11,derham14}.  These massive gravity models can address multiple problems in cosmology such as dark energy~\cite{Koivisto}, dark matter~\cite{Postnov,Mukhoyama},  inflation~\cite{Tanaka} and also in fundamental physics related to quantization of gravity~\cite{Milton}.

One generic feature of massive gravity models is that the gravitational potential has a Yukawa behavior  in the linear weak field limit, typically
parameterized as~\cite{Goldhaber74,Will97,Goldhaber10,Derham16}:
\begin{equation}
V=\frac{GM}{r}\exp (-r/\lambda_g),
\label{eq:yuk}
\end{equation}
 where $\lambda_g$ is the Compton wavelength of the graviton and is given by $\lambda_g \equiv \frac{h}{m_g c}$, where $m_g$ is the graviton mass.  
 
Basically there are three model-independent methods, which have been used to obtain graviton mass bounds~\cite{Derham16}.  The first method involves looking for a weakening of the gravitational force due to a Yukawa-like potential. The second type of constraint comes from  looking for fifth force interactions, which arise in massive gravity models.  The third type of limit  comes from the propagation of gravitational waves,  either  due to  modified dispersion relations or from difference in arrival times between gravitational waves and other astrophysical messengers (photons, neutrinos). In the gravitational wave literature, the limits from the first two types of measurements are referred to   as ``static'' bounds, whereas the limits from gravitational wave observations are referred to as ``dynamical'' bounds. In addition to these three traditional methods, one can also obtain bounds on graviton mass by studying its  implications for cosmology, such as large scale structure and late time evolution. But these are strongly model-dependent. In addition,  one can also get constraints on Yukawa gravity from the weak-field limit of certain modified gravity theories such as $f(R)$  gravity or Moffat's Scalar-Vector-Tensor gravity~\cite{demartino1,demartino2}. 
 A comprehensive summary of all the  observational/experimental bounds on the mass of the graviton  as well as future prospects can be found in  Ref.~\cite{Derham16} and a tabular summary can be found in Table 1 of the same paper. We now briefly recap the limits from these different types of methods.
 

Two years ago there was a watershed event in the history of physics, due to the direct detection of gravitational waves from the two LIGO detectors in Hanford and Livingston~\cite{LIGO1}.
These observations enabled us to obtain the most stringent \emph{dynamical} bounds on the graviton mass, by looking for dispersion in the observed signal as it propagated towards the detectors. The first detection from GW150914~\cite{LIGO1}  provided a limit of  $\lambda_g > 10^{13}$~km, or $m_g<10^{-22}$ eV~\cite{LIGO2}, based on looking for a modified dispersion relation for a non-zero graviton mass. Subsequently, a more stringent bound of $m_g<7.7 \times 10^{-23}$ eV   has been obtained using GW170104~\cite{LIGO3}. We note however that Deser~\cite{Deser16} has pointed out that  no strong field generation of radiation in massive gravity models can reproduce the observed ringdown patterns observed in LIGO. For a gravitational wave  source in our galaxy, one could also constrain this mass using the line of sight Shapiro delay from the source of the gravitational wave~\cite{Kahya}. Most recently, the direct detection of photons in coincidence with the gravitational waves from a binary neutron star merger have enabled us to constrain the mass of the graviton to $m_g \lesssim 10^{-22}$ eV~\cite{Baker17}. In the future, eLISA could obtain bounds of $m_g<10^{-26}$ eV~\cite{Will97},  and from the detection of inflationary gravitational waves from stage IV CMB experiments, one could get a bound of $m_g<3 \times 10^{-29}$ eV~\cite{cmbs4}.

Many massive gravity models give rise to a fifth force.  However, these results are theory dependent and in particular depend on how the non-linear Vainshtein mechanism operates in these theories~\cite{Derham16}. However, one common feature in these models is the existence of a Galileon-like scalar.  As of now, the bounds  on graviton mass in this category have been obtained from the decoupling limit of DGP~\cite{DGP} and dRGT~\cite{dRGT} theories. Data from lunar laser ranging experiments  give a mass bound of $m_g < 10^{-30}$ eV within the context of the decoupling limit of dRGT theory~\cite{Derham16}. From the corresponding decoupling limit of DGP, future surveys on galaxy-galaxy lensing could set a bound upto $m_g< 10^{-33}$ eV~\cite{Wyman}.

We finally recap the limits on graviton mass by looking for Yukawa-type fall off of the gravitational force. The first such bound   was obtained by Hare~\cite{Hare}, by assuming that the gravitational force from the center of the galaxy is diminished by factor of less than $\frac{1}{e}$. From this argument, a mass bound of  $m_g<6.7\times10^{-28}$ eV was obtained~\cite{Hare}. A similar reasoning was then extended by Goldhaber and Nieto to extragalactic observations of galaxy clusters~\cite{Goldhaber74}.

The current  best  limit (from all the three types of methods) on the mass of a  graviton comes from the measurements of weak lensing cosmic shear~\cite{lensing}, obtained  by comparing the variance of the modified shear convergence power spectrum in massive gravity models to the observed data~\cite{Choudhury}.\footnote{We note that the paper~\cite{Choudhury} incorrectly states that the lensing signal in  Ref.~\cite{lensing} is from a  cluster of stars at around an average redshift of z = 1.2.} By imposing the condition that the observed deviations from the $\Lambda$CDM power spectrum  are less than $1\sigma$,  a limit of  $m_g < 6\times 10^{-32}$ eV was obtained~\cite{Choudhury}. One assumption however made in obtaining this limit is that the graviton mass has no effect on the cosmological expansion, growth of structure and also the CDM transfer function. Furthermore, there is also a degeneracy in the modified power spectrum between a non-zero graviton mass and other cosmological parameters.  To evade this degeneracy, the other parameters were determined using the values of the power spectrum at smaller
values of the radius, for which the effect of a non-zero graviton mass is assumed to be negligible. These fitted parameters were then used for the limit on graviton mass using the measurements for larger values of the radius~\cite{Choudhury}.

The constraint on graviton mass in Ref.~\cite{Goldhaber74} using galaxy clusters was obtained by assuming that the orbits of galaxies in clusters are bound as well as closed and using the fact that the maximum  separation  between  galaxies from the Holmberg galaxy cluster catalog  is about 580~kpc~\cite{Holmberg}. The limit on graviton mass was obtained by positing  $e^{-1} \le \exp(-\mu_g r)$, (where $\mu_g$ is the  reciprocal of the reduced Compton wavelength)  and assuming $r=580$ kpc. This condition implies    that there are  at most $\mathcal{O}(1)$  departures from Newtonian gravity at the edge  of the galaxy cluster.
The estimated limit on graviton mass thus obtained  was   $\mu_g < 5.6 \times 10^{-25} \rm{cm^{-1}}$
or $m_g < 1.1 \times 10^{-29}$ eV or $\lambda_g > 10^{20}$~km.  We note that this is a very   rough estimate. This limit does not use any dynamical mass information for the galaxy cluster or any \emph{ansatz} for the potentials of the   different cluster components (gas, galaxies, dark matter). Also, no confidence interval was given for this upper limit. 

Furthermore, very recently it has been shown that Newtonian gravity is not the only central force that gives rise to bound orbits and one can also get bound orbits for potentials which do not satisfy Bertrand's theorem~\cite{Kazanas2,Mukherjee}. In particular, Mukherjee et al~\cite{Mukherjee} have shown that one can get single-particle bound orbits in a Yukawa potential for certain values. Thus, the main edifice upon which this bound  of $m_g < 10^{-29}$ eV has been obtained~\cite{Goldhaber74} is no longer  valid.
 
Galaxy clusters are the most massive gravitationally bound objects in the universe and provide an excellent laboratory for studying a diverse range of topics from galaxy evolution to cosmology (For reviews, see ~\cite{Voit,Kravtsov,Allen,Vikhlinin}). In the past two decades a large number of galaxy cluster surveys  in the optical~\cite{DESc,SDSSc,KIDSc,HSCc}, microwave~\cite{Bleem,ACT,Planck}, and X-ray~\cite{BCS,MACS,Reflex,XCS,Chandra,Pierre}  have come online. These surveys have enabled the discovery of a large number of galaxy clusters up to very high redshifts, allowing us to probe a  wide range of questions in astrophysics and cosmology.

Multiple observables  from these surveys such as galaxy cluster counts, gas mass fraction,  and dynamics of galaxies within clusters  have been widely used to design tests and constrain a large class of  modified theories of gravity, which dispense with dark energy~\cite{Rapetti,Raveri,Bocquet,Amendola,Vikhlinin,Lombriser,Terukina,Cataneo,Wilcox,Li}. Galaxy clusters have also been used to constrain modified gravity theories,
 which dispense with   dark matter,  e.g., various incarnations of MOND-like theories, Verlinde's entropic gravity, Moffat's MOG theory, nonlocal gravity~\cite{Silk,Rahvar,Moffat,Ettori,Moffat13,Hodson17,Finsler,Natarajan,Sanders}. However, despite the wealth of exquisite multi-wavelength galaxy cluster observations, no improvement to the initial estimate on graviton mass using galaxy clusters has been obtained after Goldhaber and Nieto's 1974 paper.\footnote{In fact this paper has not obtained any citations from any other galaxy cluster or cosmology paper, according to the ADS database. The only citations to this paper are from the gravitational wave literature or review papers, which constrain the mass of the graviton.} The only related result   is by  De Martino and Laurentis, who obtained a constraint on a variant of the Yukawa gravitational potential considered here (from the post-Newtonian limit of $f(R)$ gravity), using the thermal SZE profile of the Coma cluster from the 2013 Planck observations~\cite{demartino1,demartino2}.  In principle however, the analysis in this work could be extended to obtain a limit on the graviton mass.

Therefore, to the best of our knowledge, we are not aware of  any direct constraint  on graviton mass   using galaxy clusters  from  completed stage II or ongoing stage III dark energy experiments, or any forecast on the estimated sensitivity to graviton mass from upcoming stage IV experiments such as LSST~\cite{LSST}, Euclid~\cite{Euclid}, WFIRST~\cite{WFIRST}, etc. This is  despite the fact that one of the key science driver for these upcoming surveys is to test modified gravity theories~\cite{Jain}.

Therefore, to rectify the situation and to see how sensitive current galaxy cluster data is to graviton mass compared to very rough estimates  from four decades ago, we do a first end-to-end study to obtain a limit on graviton mass using  the Abell 1689 galaxy cluster, for which exquisite multi-wavelength data is available, allowing the reconstruction of  detailed mass models for this cluster in literature.

This manuscript is organized as follows. We discuss the dynamical modeling and mass estimates in Section~\ref{sec:dyn}. Our analysis and results can be found in Section~\ref{sec:results}. We examine the robustness of our limit to different  mass models in Section~\ref{sec:sys}.
We conclude in Section~\ref{sec:conclusions}.

\section{Dynamical Modeling of A1689}
\label{sec:dyn}
We use the galaxy cluster Abell 1689 (hereafter A1689) for our analysis. A1689 is one of the largest and most massive galaxy cluster located at a redshift of 0.18. In the past  decade, it has been subjected to intensive dynamical modeling within the context of  the $\Lambda$CDM cosmological paradigm,  using multi-wavelength observations from weak and strong lensing, SZE  and X-Ray observations~\cite{Limousin,Umetsu08,Umetsu15,Eckert} (and references therein). These observations have enabled us  to obtain estimates separately for the dark matter, gas, and galaxy components for this cluster. Most recently,  this cluster was extensively studied to see if its available data is compatible with MOND-like theories, which provide a solution for the dark matter problem from a modification of  Newtonian gravity and without the need for dark matter~\cite{Moffat16,Nieu,Hodson}. We use the same modeling from these papers to test to what extent the data is viable  with a Yukawa potential in order to constrain the graviton mass. The first step in this procedure  involves  estimating the total mass of the different components of the galaxy cluster, viz. its dark matter content, galaxies, and gas in the intra-cluster medium. We follow the same procedure
as in Ref.~\cite{Hodson}.

The total dark matter mass  can be obtained by assuming the density obeys  the Navarro-Frenk-White profile~\cite{NFW} and is given by~\cite{Hodson} : 
\begin{equation}
M_{dm} = 4\pi \rho_s r_s^3 \left[ \log\left(\frac{r_s+r}{r_s}\right)-\frac{r}{r_s+r}\right],
\label{eq:mdm}
\end{equation}
\noindent where $r_s$ and $\rho_s$ represent the dark matter halo  scale radius and scale density respectively. They are usually obtained
from the relation between the NFW concentration parameter ($c_{200}$) and the total mass at a radius 200 times the critical universe density ($M_{200}$). To calculate the total dark matter mass, we use  the NFW concentration parameters for this cluster measured by  Umetsu et al~\cite{Umetsu15}, viz. $c_{200}=10.1 \pm 0.82$, $M_{200}= (1.32 \pm 0.09 )\times 10^{15} M_{\odot} h^{-1}$, obtained  using a combination of weak and strong lensing observations. Masses obtained from weak or strong lensing do not depend on the dynamical state of the cluster and hence do not rely  on assumptions of hydrostatic equilibrium. We however note that spherical symmetry has been assumed in the dynamical mass modeling, whereas there is observational evidence that this cluster has triaxial symmetry~\cite{Umetsu15}. Although this cluster has been modeled using ellipsoidal halo~\cite{Umetsu15}, for this work spherical symmetry has been assumed throughout.

The central galaxy  (often called BCG, which is an acronym for  Brightest Cluster Galaxy) mass distribution is modeled by positing a density distribution of the form~\cite{Limousin,Nieu}:
\begin{equation}
\rho_{gal} (r) = \frac{M_{cg} (R_{co} + R_{cg})}{2\pi^2(r^2+R_{co}^2)(r^2+R_{cg}^2)}, 
\label{eq:gal}
\end{equation}
\noindent where  $M_{cg}$  and $R_{cg}$ represent the BCG mass and core radius respectively; $R_{co}$ represents the core size.
The values for  $M_{cg}$, $R_{cg}$, and  $R_{co}$ that we use for our analysis can be found in  Refs.~\cite{Moffat16,Nieu,Hodson}, which we use for our analysis.  
The gas mass is obtained using a cored Sersic profile and given by~\cite{Nieu09,Nieu}:
\begin{equation}
\rho_{gas} = 1.167m_p n_{e0} \exp\left\{k_g - k_g \left(1+ \frac{r^2}{R_g^2}\right)^{1/(2n_g)} \right\},
\label{eq:gas}
\end{equation}
\noindent where $m_p$ is the proton mass, $n_{e0}$ is the central electron density;  $R_g$ represents the radial extent of the gas; while $k_g$ and $n_g$ are dimensionless parameters which control the shape of the gas profile. The values for all these parameters can be found in Refs.~\cite{Hodson,Nieu}. The total baryonic mass $M_{bar}$ upto a given radius $R$ can be found by integrating the galaxy and gas density profiles from Eq.~\ref{eq:gal} and Eq.~\ref{eq:gas} : $M_{bar}= \int_{0}^{R} 4\pi [\rho_{gal}+\rho_{gas}] r^2 dr$. We note that these mass estimates  have been made by positing spherical symmetry.

Once, we have calculated the mass of the different components, the total acceleration assuming only Newtonian Gravity ($a_{newt}$)  from the center of the galaxy cluster is given by 
\begin{equation}
a_{newt} = G(M_{dm}+M_{bar})/R^2 . 
\label{eq:acc}
\end{equation}

\section{Results}
\label{sec:results}
In order to test the viability of Yukawa gravity, the gravitational acceleration can be obtained from the derivative of the Yukawa potential (Eq.~\ref{eq:yuk}) and is given by~\citep{Will97}:
\begin{equation}
a_{yuk} = \frac{G(M_{dm}+M_{bar})}{R} \exp \left(\frac{-R}{\lambda_g}\right) \left(\frac{1}{\lambda_g}+ \frac{1}{R}\right).
\label{eq:accyuk}
\end{equation}

In the limit that $m_g \rightarrow 0$, Eq.~\ref{eq:accyuk} will asymptote to Eq.~\ref{eq:acc}. In order to test for the  validity of this modified acceleration law, we assume that the total mass is the same as that in Newtonian gravity and we only look for deviations 
compared to ordinary gravity as a function of distance from the cluster center. This is similar to the approaches used to constrain modified theories of gravity, which dispense with the dark matter paradigm~\cite{Hodson,Nieu,Moffat16}. 

For this analysis, the NFW density profile,  the BCG density profile, and gas density estimates from literature, used to calculate the total mass have been obtained after positing a Newtonian potential. Strictly speaking, the total estimated mass would be larger within the  context of  Yukawa gravity, because of the weakness of gravity in a Yukawa potential compared to the corresponding Newtonian one.
However, a self-consistent constraint on the graviton mass is  out of the scope of the current paper and at this time would require significant additional work from the community, since one would need to determine three unknown density profiles  (the dark matter, gas, galaxy) in addition to
the graviton mass from the observational data. For the dark matter component, one would need to do a suite of $N$-body simulations in Yukawa gravity as a function of graviton mass and then obtain a parametric estimate as a function of graviton mass. Similarly, the baryonic mass would need to be determined assuming hydrostatic equilibrium in such a modified potential. Therefore, because of the large number of  additional free parameters,  doing the whole problem self-consistently in order to obtain more
robust bounds on graviton mass would pose formidable challenges.

However, if the graviton mass is very small, the deviations in the total mass estimates should  be small compared to those obtained using Newtonian gravity and the errors in our estimate of graviton mass should be negligible. Furthermore, since we shall only be interested in deviations in the acceleration profile (compared to  a Newtonian potential) for the same mass, the total mass would only be a normalization constant and would not make a  difference to the final limit. Therefore, similar to what is usually done in  constraining alternate gravity theories, which dispense with the dark matter paradigm (see e.g. ~\cite{Hodson} and references therein), we assume that the total density profile is the same  as in Newtonian gravity and then look for deviations in the acceleration profile  as a function of distance from the center of the cluster to constrain departures from a standard Newtonian acceleration profile. Due to the above assumption, the limit is of course conservative. In Sect.~\ref{sec:sys}, we shall examine how the limit on graviton mass changes when varying the mass models for the cluster used here.  

We also point out that if we posit that dark matter is made up of  massive gravitons (see e.g.~\cite{Postnov}), then only the dark matter potential would be modified while  the other terms in the  potential would be unchanged and our limits would be different. Here, we assume that all the distinct mass  components (gas, galaxy, dark matter) uniformly obey the Yukawa potential and  dark matter is some hypothetical elementary particle, with the same gravitational laws as the baryonic components.

To quantify the deviations between Newtonian and Yukawa gravity as a function of distance from the center of the cluster, we construct a $\chi^2$ functional given by,
\begin{equation}
\chi^2= \sum\limits_{i=1}^N \left(\frac{a_{newt}-a_{yuk}}{\sigma_a}\right)^2,
\label{eq:chi}
\end{equation}
\noindent where $a_{newt}$ and $a_{yuk}$ are  given by Eqs.~\ref{eq:acc} and ~\ref{eq:accyuk}; $\sigma_{a}$ is the uncertainty in the estimated acceleration. To get the 90\% c.l. upper limit on the mass of the graviton, we find the threshold value of $m_g$ for which $\Delta \chi^2>2.71$~\cite{NR}, where $\Delta \chi^2=\chi^2 -\chi^2_{min}$. We note that $\chi^2_{min}$=0 corresponds to a zero graviton mass. Therefore, $\Delta \chi^2$ is identically equal to $\chi^2$ from Eq.~\ref{eq:chi}. Since, the mass of the graviton cannot be negative, $m_g=0$  is a physical boundary. Therefore, in such cases $\Delta \chi^2$ values for a given confidence interval could in principle get modified compared to the values in Ref.~\cite{NR}. To obtain the modified $\Delta \chi^2$, we use the procedure recommended  by the Particle Data Group~\cite{PDG,Messier}, which has  previously been used  for neutrino oscillation analysis~\cite{Superk}. The effect of the physical boundary is determined by the difference between the minimum value of $\chi^2$
and the value of $\chi^2$ at the boundary of the physical region. If the minimum value of $\chi^2$ occurs at the physical boundary (which is true in our case), then $\Delta \chi^2$  intervals for a given confidence interval are the same  as without a physical boundary~\cite{Messier}. Therefore, to obtain the 90\% confidence level upper limit  we  choose $\Delta \chi^2=2.71$, which is the same as the value without a physical boundary.
Alternately, the modified $\Delta \chi^2$ threshold can also be obtained using the Feldman-Cousins method~\cite{Feldman}, which requires extensive Monte-Carlo simulations. However, they have been shown to not differ too much compared to the method use here~\cite{Messier}.

\begin{figure}
\centering
\includegraphics[width=0.5\textwidth]{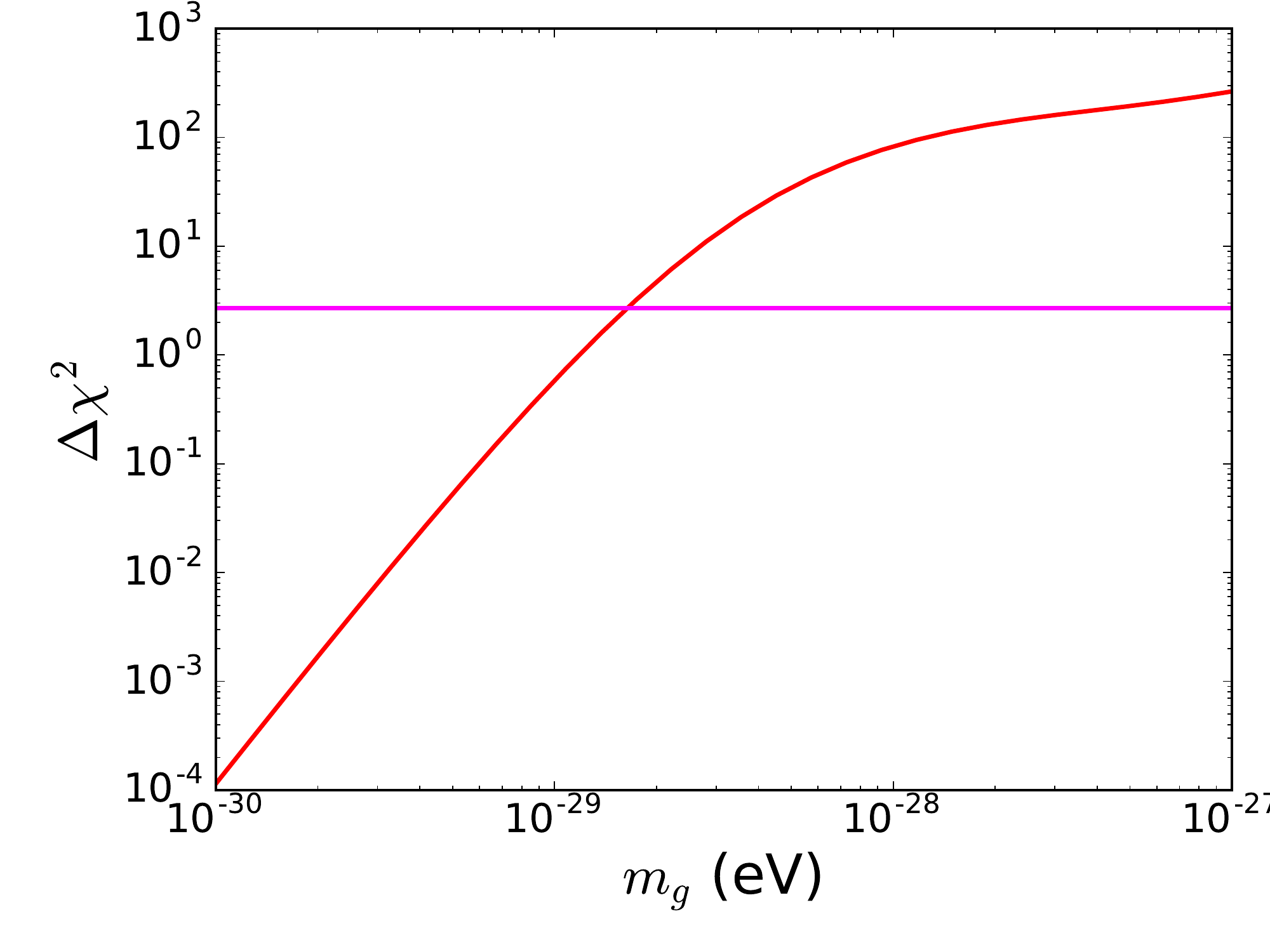}
\caption{$\Delta \chi^2$  as a function of graviton mass. The horizontal line at $\Delta \chi^2 =2.71$ gives the 90\% c.l. upper limit  on graviton mass of $1.37 \times 10^{-29}$ eV  or a lower limit on the Compton wavelength of $\lambda_g > 9.1 \times 10^{19}$ km. We note that  since $\chi^2_{min}=0$ for $m_g=0$, this is mathematically equivalent to the $\chi^2$ functional defined in Eq.~\ref{eq:chi}.}
\label{fig1}
\end{figure}

\begin{figure}
\centering
\includegraphics[width=0.5\textwidth]{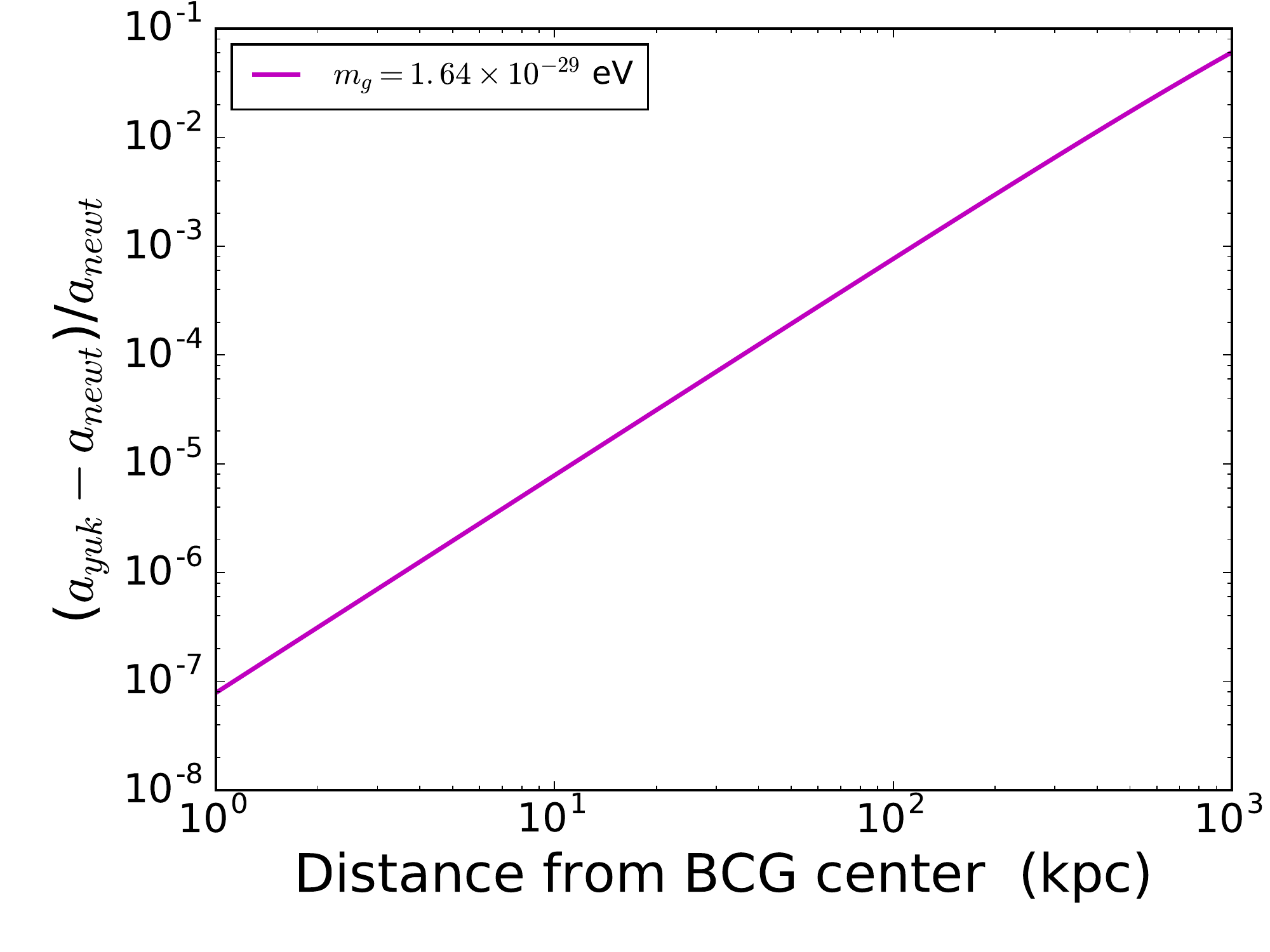}
\caption{Fractional absolute deviation between acceleration computed assuming Yukawa gravity (for a graviton mass of $m_g=1.37 \times 10^{-29}$ eV, corresponding to the 90\% cl upper limit) as a function of distance from the center of the central galaxy of the cluster (usually referred to as BCG). The fractional deviation is about 10\% at 1 Mpc.}
\label{fig2}
\end{figure}

 We calculated the $\Delta \chi^2$ for 24 points between roughly 1 and 3000 kpc, for which
errors in acceleration have been estimated from existing observations~\cite{Nieu}, for which spherical symmetry has been assumed. 
The first 12 points were located at radii between 3 and 271 kpc, for which the errors in acceleration have been estimated from the line of sight mass density~\cite{Nieu09}, obtained using strong lensing observations~\cite{Limousin}.
The remaining 12 data points were distributed between 125~kpc and 3~Mpc and the errors in acceleration were estimated  from the weak lensing shear profiles~\cite{Umetsu15}. We note that the errors in acceleration data do not include any errors in determination of the radii. $\chi^2$ was then estimated from Eq.~\ref{eq:chi} for these 24 data points by calculating $a_{newt}$ and $a_{yuk}$ at these radii and using the errors in acceleration estimated in Ref.~\cite{Nieu}. This plot is shown as a function of graviton mass in Fig.~\ref{fig1}.
The 90\% c.l. upper limit on the  mass of a graviton obtained  from $\Delta \chi^2=2.71$, is given by $m_g<1.37 \times 10^{-29}$ eV, corresponding to a  Compton wavelength of  $\lambda_g >9.1 \times 10^{19}$ km. For this value of mass, we also show the fractional deviation between the ordinary Newtonian acceleration and that assuming the Yukawa potential in Fig.~\ref{fig2}. We can see that for this graviton mass, the differences are less than 1\% upto 200 kpc and about 10\% at about 1 Mpc.

\section{Effect of systematic errors}
\label{sec:sys}

We now examine  the sensitivity of our limit on graviton mass to different mass models for the three components of A1689, compared to the results in the previous section. A tabular summary of all these
upper limits on varying the mass models can be found in Table~\ref{tab1}.

We  start with the dark matter part. A large number of groups have obtained different NFW parameters for this cluster using weak and strong lensing data. (See Table 9 of Ref.~\cite{Umetsu15}). We first examine how our result changes with different NFW parameters from the literature by 
considering two values of the concentration parameter, which span the full range of the  estimated values and for which spherical symmetry is assumed.  The lowest  value for $c_{200}$ for this cluster corresponds to  $c_{200}=5.71$ for $M_{200}=1.81 \times 10^{15} M_{\odot} h^{-1}$~\cite{Morandi}. The corresponding upper limit on graviton mass is at $1.18 \times 10^{-29}$ eV. At the other extreme, when we choose $c_{200}=12.2$  for $M_{200}=0.83 \times 10^{15} M_{\odot}$, we get $m_g <1.43 \times 10^{-29}$ eV.

Even though, most of the mass in this galaxy cluster is made up  of  dark matter, we don't expect any major changes with different BCG or gas mass profiles. Nevertheless, to check this, instead of  Eq.~\ref{eq:gal}, we used the Hernquist profile~\cite{Hernquist} (similar to Ref.~\cite{Hodson17}) for the BCG mass with the same parameters as in Ref.~\cite{Hodson17}. With this new galaxy mass profile, the new graviton mass limit is the same as before.

Therefore, the change in the limit on the graviton mass by varying our \emph{ansatz} for the  mass models is less than 15\% and does not change the ballpark estimate on the limit on graviton mass of $m_g \lesssim 10^{-29}$ eV.

\begin{table}
\begin{tabular}[t]{|c|c|c|c|c|} \hline
$M_{200}$ & $c_{200}$ & Gas Mass   & BCG Mass  & $m_g ( 10^{-29} eV)$  \\ 
($10^{15} M_{\odot} h^{-1}$)  & & & \\
\hline
1.32 & 10.1~\cite{Umetsu15}  & Eq~\ref{eq:gas} & Eq~\ref{eq:gal}  & $<1.37$ \\
1.81 &  5.7~\cite{Morandi} &Eq~\ref{eq:gas}   & Eq~\ref{eq:gal} & $<1.18 $ \\
0.83 & 12.2~\cite{Corless} & Eq~\ref{eq:gas} & Eq~\ref{eq:gal} & $<1.43$ \\
1.32 & 10.1~\cite{Umetsu15}  & Eq~\ref{eq:gas} & Hernquist &  $<1.37$ \\
\hline
 
\end{tabular}
\caption{Sensitivity of the graviton mass limit to different models of dark matter potential (different NFW fits), gas mass, and BCG mass. The first two columns indicate $c_{200}$ and $M_{200}$ values used for the NFW profile to calculate the total dark matter mass from Eq.~\ref{eq:mdm}. The next two columns indicate the corresponding equation (or profile), from which the gas   and BCG mass was estimated. The final column indicates the upper limit on graviton mass (expressed as a multiple of $10^{-29}$ eV.)}
\label{tab1}
\end{table}

\section{Conclusions}
\label{sec:conclusions}

In 1974, a limit on  graviton mass  of $m_g<1.1 \times 10^{-29}$ eV was obtained from galaxy clusters, using the fact that the orbits of galaxy clusters are bound up to 580 kpc~\cite{Goldhaber74} and such closed bound orbits can only exist within Newtonian gravity. However, recently it has been shown that one can get closed bound orbits for a   Yukawa potential~\cite{Mukherjee}. Therefore, the main premise used to obtain the mass bound limit from galaxy clusters in Ref.~\cite{Goldhaber74} can no longer be justified and this result should no longer be quoted in the literature.

Subsequently, even though  a huge amount of work has been done  in testing a plethora of  modified gravity theories with  galaxy clusters using optical, X-ray, and SZE data, we are not aware of any other work on estimating a bound on graviton mass from clusters, despite a wealth of new precise observational data  in the past decade, courtesy a whole slew of   multi-wavelength surveys. 

We obtain a limit on graviton mass from A1689 using an independent method compared to Ref.~\cite{Goldhaber74}. We use recent dynamical mass models of the different components of galaxy cluster A1689, obtained using X-ray, weak and strong  lensing data~\cite{Nieu,Moffat16,Hodson} to obtain a limit on the graviton mass. 
For this purpose, we assume that the potential due to the gas, galaxy, and dark matter all follow a Yukawa behavior, due to non-zero graviton mass. We then look for deviations from the estimated acceleration data (assuming validity of Newtonian gravity) and a Yukawa potential,   and find the critical graviton mass for which the $\Delta \chi^2$  difference between the two potentials crosses 2.71. This gives us a 90\% c.l. upper  bound on the graviton mass of $m_g<1.37 \times 10^{-29}$ eV or on  the Compton wavelength $\lambda_g > 9.1 \times 10^{19}$ km. We also checked how the limit varies with different mass models for the dark matter and BCG potential. We find that  the maximum variation in the limit on graviton mass is about 15\%
and thus does not change the ballpark estimate of our limit, which is $\mathcal{O}(10^{-29})$ eV.

We should point out that  the fact that our  upper limit is approximately of  the same order 
of magnitude as that obtained by Goldhaber and Nieto~\cite{Goldhaber74} is only a coincidence. The maximum size they  assumed for the  galaxy cluster orbits is  about 580 kpc, as this was the size of  the  largest known clusters in 1974. Using this estimate for the  size, they obtained an upper limit of $\mathcal{O} (10^{-29})$ eV, which is of the  same order of magnitude as ours.
In principle, one could trivially apply the same method~\cite{Goldhaber74} to some of the galaxy superclusters currently known. For example, the recently discovered Saraswati supercluster~\cite{Bagchi} (whose size is at least 200 Mpc) would  yield a more stringent upper limit on the graviton mass of about $3 \times 10^{-32}$ eV. However, as mentioned earlier the underlying assumptions behind this argument used to obtain the limit are incorrect.

We however note that to obtain our limit, mass estimates for the different components (dark matter, gas, galaxy) have been obtained assuming Newtonian gravity, since otherwise the whole problem  of simultaneously  determining the mass of the three unknown components in addition to the graviton mass becomes currently intractable, given the large number of free parameters. However, in this case  since the limit has 
been obtained by using deviations from Newtonian acceleration profile as a function of the distance from the galaxy cluster center and the total mass of each component would mainly act as a normalization constant and  not make a big difference to our final limit.

Given the large number of upcoming Stage IV  dark energy experiments  such as LSST~\cite{LSST}, Euclid~\cite{Euclid}, WFIRST~\cite{WFIRST} etc, it would be interesting to  estimate the expected improvement in the limit on graviton mass compared to the result obtained in this work. We  now carry out an order of magnitude estimate of the same.

Some of our errors in acceleration data (at radii less than about 150 kpc) come from strong lensing measurements~\cite{Limousin}, which use Hubble Space Telescope data. Therefore, we do not expect significant improvements in the strong lensing based error estimates. The acceleration errors at higher radii are obtained from weak lensing measurements using Suprime-Cam data~\cite{Umetsu15}. 
For Euclid and other stage IV experiments, the multiplicative bias from the shear must be less than 0.1\%~\cite{Amara,Euclid}.
If we evaluate the acceleration errors from these predicted shear errors for distances  greater than 150 kpc and combine it with current errors from strong lensing for smaller radii, we expect a 90\% confidence upper limit of $m_g<2.75 \times 10^{-30}$ eV. This is still not as sensitive as the current best limit on graviton mass~\cite{Choudhury}. However, we caution that this is only a ballpark estimate of expected improvement sensitivity. More detailed forecasting studies need to be done by the relevant working groups from the various stage IV dark energy experiments. One key missing ingredient
needed for that purpose is the generation of   $N$-body simulations in Yukawa gravity and the calculation of  the corresponding halo mass functions.

\begin{acknowledgements}
We are thankful to the anonymous referee for detailed critical feedback on the manuscript.
We would like to thank Sabine Hossenfelder for explaining the recent resurgence of interest in massive gravity theories. 
We are grateful to Theo Nieuwenhuizen for providing us the data in ~\cite{Nieu} and for comments on the draft. We  also acknowledge    Ivan De Martino, Fred Goldhaber, Alistair Hodson,  and Mark Messier   for useful correspondence and discussions.

\end{acknowledgements}

\bibliography{graviton}
\end{document}